%% file: main.tex
\documentclass[twocolumn,reprint,superscriptaddress,superscriptaffil,prb,dblfloatfix,floatfix, amsmath,amssymb]{revtex4-2} 

\usepackage{soul}
\usepackage{multirow}
\usepackage{float}
\usepackage{amsmath}
\usepackage{xfrac}
\usepackage{caption}
\usepackage{subcaption}
\usepackage{tabularx}
\usepackage{ragged2e} 
\usepackage[dvipsnames]{xcolor} 
\usepackage[normalem]{ulem} 
\usepackage[bottom]{footmisc}
\usepackage[colorlinks=true, allcolors=blue]{hyperref}
\usepackage{graphicx}
\usepackage{dcolumn}
\usepackage{bm}
\usepackage{pdfpages} 

\include{math_HP}
\include{math_DFT}

\makeatletter
\AtBeginDocument{\let\LS@rot\@undefined}
\makeatother

\begin{document}
\title{Revealing Local Field Effects at the Electrical Double Layer with Efficient Open Boundary Simulations under Potential Control}
\author{Margherita Buraschi}
\email{m.buraschi20@imperial.ac.uk}
\affiliation{Department of Chemistry, Imperial College London, White City Campus, London W12 0BZ, UK}%
\author{Andrew P. Horsfield}
\affiliation{Department of Materials and Thomas Young Centre, Imperial College London, South Kensington Campus, London SW7 2AZ, UK}%
\author{Clotilde S. Cucinotta}
\email{c.cucinotta@imperial.ac.uk}
\affiliation{Department of Chemistry and Thomas Young Centre, Imperial College London, White City Campus, London W12 0BZ, UK}%

\date{\today}

\begin{abstract}
\input{abstract}
\end{abstract}

\maketitle

\input{section_1}
\input{section_2}

\input{section_3}

\input{conclusions}
\input{acknowledgements}
\bibliography{bibliography}

\clearpage



\section*{Supplementary material}

# Supporting Material

Margherita Buraschi[1], Andrew P. Horsfield [2], and Clotilde S. Cucinotta[3]

[1]Department of Chemistry, Imperial College London, White City Campus, London W12 0BZ, UK
[2]Department of Materials and Thomas Young Centre, Imperial College London, South Kensington Campus, London SW7 2AZ, UK
[3]Department of Chemistry and Thomas Young Centre, Imperial College London, White City Campus, London W12 0BZ, UK

November 3, 2023

# 1 Code efficiency

Profiling of the code shows that the HP-DFT formalism did not result in a relevant slowdown of the SCF cycle compared to standard DFT algorithms. For example, the average time per SCF cycle step for a system with 3904 electrons was circa 5.7 s for standard DFT, while the HP-DFT calculation at $\Delta\mu = 0$ eV took around 6.6 s per cycle. Even at $\Delta\mu = 1$ eV the average time per SCF cycle step was around 6.8 s. This minimal time difference can be attributed to the fact that the necessary quantities to determine the HP-DFT occupation numbers, $f_i$, using Equation (3) of the paper are already computed by the underlying DFT code. As a result, the HP-DFT approach proves to be computationally efficient, making it particularly appealing to control and tune the electrochemical potential in a simulation. It is worth noting that this efficiency extends to applications in combination with *ab initio* molecular dynamics (AIMD) calculations; although ongoing preliminary tests show a mild slowdown in a system with a large number of molecules in solution, the average time per SCF cycle remains comparable to that of a standard AIMD calculation.

# 2 Single water molecule adsorbed on a charged surface

Initially, we performed a set of calculations with only a single water molecule adsorbed on both the internal surfaces of the Pt(111)(6×6×3) capacitor. We performed HP-DFT calculations at $\Delta\mu = 0, 1$ and 4 eV and we studied the response of the adsorbates to the surface charge redistribution induced by the application of a potential.

Figure (S1a) shows the charge distribution as a function of the applied $\Delta\mu$. As expected, the charge of the plates varies only on the internal surfaces of the capacitor; at the same time, the molecule adsorbed on the left surface becomes more positive while the one adsorbed on the right surface becomes more negative. Figure (S1b) shows water re-orientation as a response

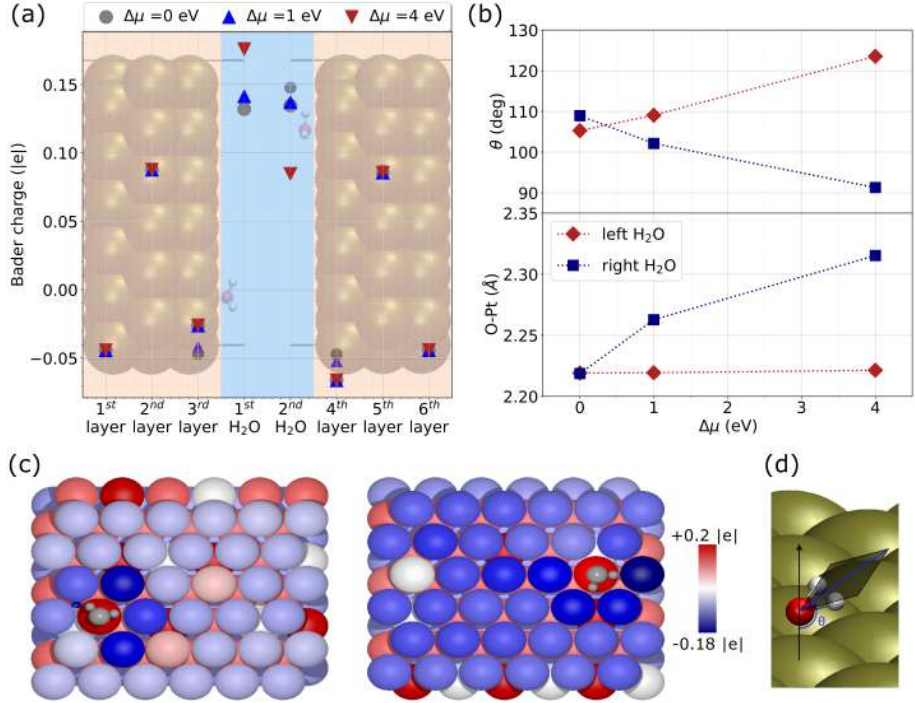

Figure S1: Pt(111)(6x6x3) system plus water molecule: (a) average excess Bader charges per atom in each layer of the plates and per molecule; (b) $\theta$ and O-Pt distance for the molecules adsorbed on the left (red) and on the right (blue) slab; (c) Excess Bader charges on the left and right surface; (d) definition of $\theta$.

to the varying surface charge. As $\Delta\mu$ increases, the O-Pt distance decreases for the molecule on the left surface while it increases for the molecule adsorbed on the right surface. At the same time, defining $\theta$ as the angle between the normal to the slab surface and the water dipole (shown in Figure (S1d)), we observe that its value increases on the left surface while it decreases on the right one. In other words, the water molecule adsorbed on the left slab rotates to point its oxygen towards and closer to the surface; the opposite happens on the right slab, where the water molecules rotate to move its oxygen away from the surface.

Figure (S1c) shows the charge distribution on both the left and right surface at $\Delta\mu = 4$ eV. In both cases, the Pt atom acting as the adsorption site carries a more positive charge while its direct neighbours carry, instead, a more negative charge (compared to the other surface atoms). The electronegative platinum surface withdraws charge from the water's oxygen, resulting in a positively charged molecule. The electron charge originally on water is then transferred from the adsorption site to the surface atoms surrounding it, which become more negative.

# 3 Excess Bader charges on Pt(111)(6×6×3) bilayer system

| $\Delta\mu$ (eV) | Total charge on plate ($|e|$) | | Average charge on molecule ($|e|$) | |
|---|---|---|---|---|
| | Left plate | Right plate | 1$^{st}$ ads. layer | 2$^{nd}$ ads. layer |
| | | **H-down configuration** | | |
| -4.00 | -0.476 | 0.365 | 0.066 | -0.058 |
| -1.00 | -0.294 | -0.029 | 0.074 | -0.046 |
| 0.00 | -0.225 | -0.161 | 0.077 | -0.044 |
| 1.00 | -0.161 | -0.293 | 0.08 | -0.042 |
| 4.00 | 0.042 | -0.707 | 0.09 | -0.036 |
| | | **H-up configuration** | | |
| -4.00 | -0.539 | 0.312 | 0.061 | -0.034 |
| -1.00 | -0.359 | -0.259 | 0.064 | -0.009 |
| 0.00 | -0.3 | -0.459 | 0.059 | -0.001 |
| 1.00 | -0.219 | -0.595 | 0.062 | 0.006 |
| 4.00 | 0.013 | -0.998 | 0.07 | 0.011 |

Table S1: Total Bader charge on the left and right metallic plates and average Bader charge per molecule in the 1$^{st}$ and 2$^{nd}$ adsorption layer for the H-down (top) and H-up (bottom) configuration.

Table (S1) reports the total Bader charge on the metallic plates as well as the average Bader charge per molecule in the 1$^{st}$ and 2$^{nd}$ adsorption layer. As $\Delta\mu$ goes from $-4$ eV to 4 eV, we can observe the surface charge on the left plate becoming more positive; at the same time, the average charge on the molecules of both the 1$^{st}$ and the 2$^{nd}$ adsorption layer becomes more positive. This indicates that more charge is passed from the water to the surface as the plate becomes more positively charged. Noticeably, the molecules in the 2$^{nd}$ adsorption layer of the H-up configuration have a much smaller charge than their counterpart in the H-down configuration: this observation suggests that these waters are less bonded to the surface and are stabilized mainly by their hydrogen bonds with the molecules in the 1$^{st}$. The overall charge of the left plate + water bilayer is compensated by the surface charge on the right plate (the counter electrode) which, accordingly, becomes negative with increasing $\Delta\mu$.

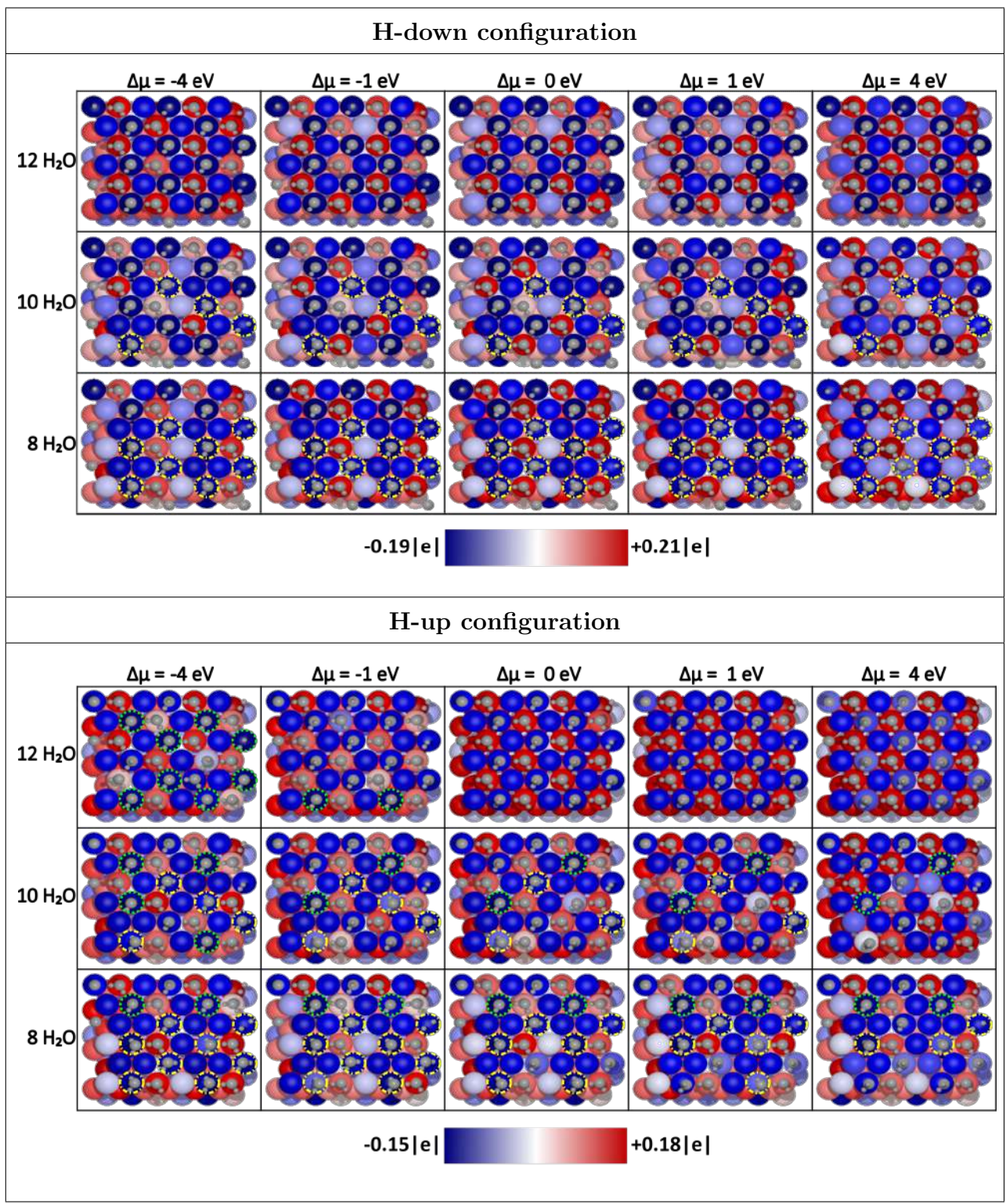

Table S2: Excess Bader charges on the left plate of the Pt(111)(6×6×3) plus H-down and H-up bilayer; both the full bilayer and the low coverage systems are shown. The green dotted circles in the H-up systems indicate the molecules in the 2$^{nd}$ adsorption layer which flip from an H-up to and H-down configuration as $\Delta\mu$ becomes more negative. The yellow dot-dashed circles in both the H-down and the H-up systems indicate the molecules in the 2$^{nd}$ adsorption layer which bring their dipole almost parallel to the surface.

Table (S2) shows the excess Bader charges on the left plate of the Pt(111)(6×6×3) plus bilayer systems in the H-down and H-up configurations. Both the full bilayer and the low coverage systems are shown. As a result of the removal of some molecules from the 1$^{st}$ layer, some molecules in the 2$^{nd}$ are not stabilized by the hydrogen bonds and adopt a configuration that neither H-up nor H-down: their dipole becomes almost parallel to the surface, with both hydrogens pointing slightly down towards the plate. In the H-down system, the number of molecules in the 2$^{nd}$ layer which become almost parallel to the surface does not seem to be particularly influenced by the value of $\Delta\mu$ (yellow dot-dashed circles in Table (S2), H-down configuration). In the H-up system, on the other hand, we observe an interesting trend. When the coverage is high, there is a progressively higher number of 2$^{nd}$ layer molecules flipping from an H-up to an H-down configuration as $\Delta\mu$ becomes more negative (green dotted circles in Table (S2), H-up configuration). In the low coverage systems, however, when the potential is low the response to polarisation occurs mostly through an increasing number of molecules in the 2$^{nd}$ layer re-orienting themselves in the almost parallel configuration (yellow dot-dashed circles in Table (S2), H-up configuration).

# 4 Pt(111)(6×6×3) bilayer system in the H-up configuration

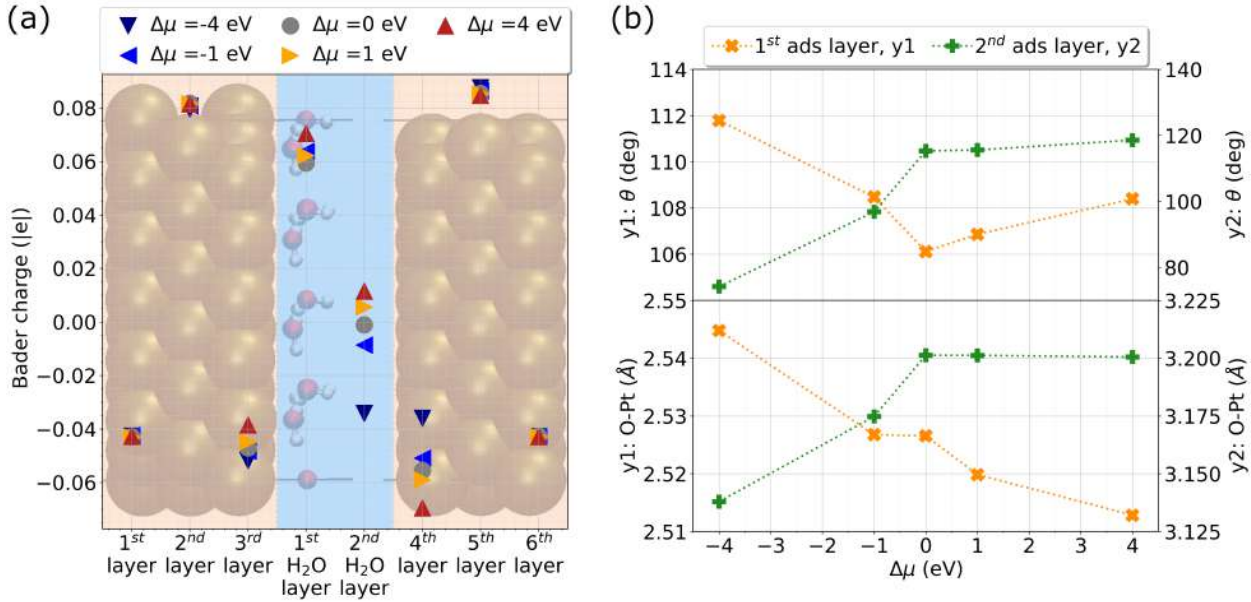

Figure S2: Pt(111)(6×6×3) system plus water H-up bilayer: (a) average excess Bader charges of each atom of the plates and on each molecule of the bilayer; (b) average $\theta$ and O-Pt distance for the molecules in the 1$^{st}$ (orange) and 2$^{nd}$ (green) adsorption layer.

Figure (S2a) shows the charge distribution as a function of the applied $\Delta\mu$ while Figure (S2b) shows water re-orientation in the 1$^{st}$ (orange) and 2$^{nd}$ (green) adsorption layer as a response to the varying surface charge. The trends discussed in the paper for charge distribution and water re-orientation are clearly visible only for positive values of $\Delta\mu$ while

5

the data for negative values of $\Delta\mu$ is more disordered (as can also be observed in Table (S1)). This happens because several molecules of the $2^{\text{nd}}$ adsorption layer flip from an H-up to an H-down configuration.



\end{document}

%% file: math_HP.tex
\newcommand{\FermiDistribution}{f^{(p)} (\epsilon_i)= \frac{1}{1+\exp[(\epsilon_i - \mu_{p})/k_bT_p]}}

\newcommand{\OccupationWeakSolutionProbes}{ f_{i} = \frac{ \alpha \overline{f}(\epsilon_{i}) \sum_s |C_{j_si}|^2  + \sum_p f^{(p)} (\epsilon_{i}) {|C_{j_pi}|}^2 } {\alpha \sum_s |C_{j_si}|^2 + \sum_s |C_{j_si}|^2 } }

\newcommand{\FermiDistributionSolution}{\overline{f}(\epsilon_i) = \frac{1}{1+\exp[(\epsilon_i - \overline{\mu})/k_bT]}}

%% file: math_DFT.tex
\newcommand{\ElectronDensity}{\rho(\textbf{r})=\sum_{nm} P_{nm}\phi_n(\textbf{r})\phi_m^{*}(\textbf{r})}

\newcommand{\DensityMatrix}{P_{nm}=\sum_i f_{i} C_{ni}C^{*}_{mi} }

%% file: abstract.tex
A major challenge in modelling interfacial processes in electrochemical (EC) devices is performing simulations at constant potential. This requires an open-boundary description of the electrons, so that they can enter and leave the computational cell. To enable realistic modelling of EC processes under potential control we have interfaced Density Functional Theory with the hairy probe method in the weak coupling limit (DOI: 10.1103/PhysRevB.97.045116). Our implementation was systematically tested using simple parallel-plate capacitor models with pristine surfaces and a single layer of adsorbed water molecules. Remarkably, our code's efficiency is comparable with a standard DFT calculation. We reveal that local field effects at the electrical double layer induced by the change of applied potential can significantly affect the energies of chemical steps in heterogeneous electrocatalysis. Our results demonstrate the importance of an explicit modelling of the applied potential in a simulation and provide an efficient tool to control this critical parameter.

%% file: section_1.tex
\section{Introduction}
The functionality of electrochemical (EC) devices is primarily determined by the electrified interface (EI) where the formation of an electrical double layer (EDL) occurs. To achieve a rational design of EC devices, an atomistic understanding of the EDL is highly desirable \cite{Hansen2016} as its equilibrium structure and response to applied potential directly affect the thermodynamics and kinetic barriers for interfacial mass and charge transfer (CT) \cite{Schmickler&Santos}. First principles computational studies are well suited for this task as they achieve electronic level resolution \cite{Khatib2021,Rossmeisl2015}. \\
The chemical reactivity and structural properties of the EDL are tuned by the electrode potential. However, including the potential in computer simulations remains a challenge \cite{Khatib2021,Rossmeisl2015}. EC systems are often studied using density functional theory (DFT) in the canonical ensemble (constant temperature, volume and number of particles). An elegant approach to incorporate the electrode potential is N\o{}rskov's computational hydrogen electrode (CHE), which has proven successful in replicating the experimentally well-known volcano plot for hydrogen evolution reaction (HER) \cite{Norksov2004,Norskov2005}. However, the potential is implicitly included as a thermodynamic correction only to the electron-proton transfer steps of the reaction. CHE fails to capture the potential-induced charging of the interface under operating conditions as it considers steps such as $\mathrm{H_2O}$, $\mathrm{O_2}$ or $\mathrm{CO_2}$ adsorption as independent of the applied potential. This has recently been proven to be incorrect \cite{Dudzinski2023}. \\
A more realistic way to explicitly introduce the effect of a potential in canonical simulations is to control the charge on the electrode surface and evaluate \textit{a posteriori} the relation between potential and charge \cite{Rossmeisl2015}. This is achieved by adding a judicious distribution of counter-charges that enforce cell neutrality \cite{FilholNeurok2006,Sprik2010,Sprik2012,Rossmeisl2013,Rossmeisl2016,Gross2016,Le2017,Gross2018,Khatib2021,Neugebauer2021,Darby2022,Cheng2023,Neugebauer2018}. Despite the undeniable insight these approaches have provided about the EIs, fixed-charge simulations remain conceptually different from fixed-potential simulations. All of the above approaches need either large samples or repeated simulations \cite{Rossmeisl2015}, fictitious distributions of charge, lack of flexibility in varying the electrode potential, and an inability to determine directly the potential dependence of transition states' charges and activation barriers. Ultimately, the electrode potential is not guaranteed to remain constant during the simulation as charge transfer at the interface causes fluctuations in the electrode charge and, consequently, in its potential.\\
Grand-Canonical DFT (GC-DFT) \cite{Mermin1965, Otani2008, Bonnet2012, Sundararaman2017} allows the number of electrons ($N_e$) in the cell to vary while keeping the chemical potential of the electrons ($\mu_e$) fixed. The grand potential of the system, rather than the free energy, is minimized as a function of the electron density \cite{Lozovoi2001}. This formulation is suitable for describing metallic slabs that form part of an electric circuit. However, GC-DFT simulations are generally harder to converge than canonical ones \cite{Lozovoi2001,Otani2008,Pasquarello2018} and often use implicit models to describe the electrolyte \cite{Bonnet2012, Andreussi2014, Otani2018, Melander2019, Hormann2019, Hille2019, Sayer2019, Ringe2022}. \\
As the setup of electrochemical cells reflects the computational framework used in molecular electronics, Non-Equilibrium Green's Function (NEGF) combined with DFT is emerging as a method to calculate the effects of bias on the electronic properties and atomic forces at the EI \cite{Hansen2016, Pedroza2018}. While this approach enables a fine-tuned control of the potential in a simulation, the high computational cost of the NEGF+DFT calculations in current implementations makes simulating large systems difficult and expensive. \\
In This paper, we propose a novel and efficient methodology to introduce electronic open-boundaries into DFT calculations. The hairy probes (HP) formalism \cite{Horsfield2016,Zauchner2018} is an open-boundaries formalism suitable for multi-terminal EC problems. Here we interface HP with the traditional Kohn-Sham DFT formalism, as implemented in CP2K \cite{Kuhne2020}. The aim is to develop a computational tool, hairy-probes DFT (HP-DFT), which enables electronic structure and force calculations under potential control.\\

%% file: section_2.tex
\section{Methodology}
\subsection{HP-DFT formalism} \label{Section:HP-DFT}
The foundation of the HP formalism is the concept of \textit{probe}. Each probe is a virtual, atomically thin lead coupled at one end to an electron reservoir, with a known electrochemical potential ($\mu_p$) and temperature ($T_p$), and at the other end to an orbital of the system. How strongly the probes are coupled to the system is defined by the coupling strength parameter $\Gamma_p$. Each probe can act as both a source and sink of electrons at a given $\mu_p$. The probes are assumed to be in equilibrium with their respective reservoir so that the electrochemical potential and temperature of the electrodes correspond to those of the reservoirs they are coupled to \cite{Zauchner2018}. The HP formalism implemented here allows for any number of probes, making it suitable for multi-terminal EC problems. 
\\
The limiting case of probes weakly coupled to the system ($\Gamma_p \rightarrow 0 $) is suitable for describing EC systems \cite{Horsfield2016} as the electric current is carried by ions in the electrolyte, whose diffusion rate sets the electron conduction rate between electrodes. Since ion diffusion is much slower than ballistic transport within the probes, it is reasonable to assume that the weakly coupled probes do not restrict current flow. \\
The probes responsible for maintaining the electrochemical potential difference across the system (\textit{main probes}) are typically coupled only to the atoms in the contact regions. This might result in the situation where atoms located far from the contacts remain uncoupled to the reservoirs, and thus not populated with electrons. Thus, we also implemented a secondary type of probe coupled even more weakly to the remaining atoms (\textit{solution probes}): they have no obvious physical meaning and they simply prevent the nonphysical situation where molecules in solution have no electrons. \\
The HP formalism is introduced into the DFT through the general representation of the electron density:
\begin{equation}
    \ElectronDensity \label{Math:electron_density} 
\end{equation}
with 
\begin{equation}
    \DensityMatrix \label{Math:density_matrix}
\end{equation}
where $\phi_n(\textbf{r})$ are the basis functions used to represent the molecular states and $P_{nm}$ is the density matrix, described by Equation (\ref{Math:density_matrix}), where $C_{ni}$ are the Molecular Orbital (MO) coefficients and $f_i$ are the occupation numbers of the molecular states. DFT codes often employ the Fermi-Dirac distribution, to define $f_i$. In the weak coupling HP formalism, the occupation numbers are instead given by:
\begin{equation}
    \OccupationWeakSolutionProbes  \label{Math:generalized_occupancy_HP_weak_solution}
\end{equation}
with
\begin{eqnarray}
    & \FermiDistribution \label{Math:fermi_distribution_HP} \\
    & \FermiDistributionSolution \label{Math:fermi_distribution_HP_solution}
\end{eqnarray}
In Equation (\ref{Math:generalized_occupancy_HP_weak_solution}), $\epsilon_i$ is the MO energy for orbital $i$, $C_{j_pi}$ is the coefficient of MO $i$ for the atomic orbital $j_p$ coupled to probe $p$, and $f^{(p)}$ is a Fermi-Dirac distribution defined by the electrochemical potential $\mu_{p}$ and electronic temperature $T_p$ of the electrons in the reservoir for probe $p$ (Equation (\ref{Math:fermi_distribution_HP})); $C_{j_Si}$ are the MO coefficients of the atoms coupled with the solution probes and $\overline{f}(\epsilon_i)$ is a Fermi-Dirac distribution defined by $\overline{\mu}$, the average electron electrochemical potential of the system (Equation (\ref{Math:fermi_distribution_HP_solution})). The way we define $\mu_{p}$ is $\mu_{p} = \overline{\mu} + \Delta \mu_p$.  Finally, $\alpha$ is a small parameter that allows smooth switching of the electron population between those from the main probes and that for the solution probes. For a full derivation of these formulas, we direct the reader to references \cite{Horsfield2016} and \cite{Zauchner2018}. 
\subsection{HP-DFT implementation and model systems}
To implement the HP-DFT formalism we developed and included in the QUICKSTEP \cite{Hutter2005_QS} module of CP2K \cite{Kuhne2020} a module to calculate $f_i$ using Equation (\ref{Math:generalized_occupancy_HP_weak_solution}). \\
We evaluated our HP-DFT implementation using parallel-plate capacitor models, consisting of two Pt(111) slabs separated by 10 $\mathrm{\AA}$. We tested models with varying surface areas and slab thicknesses, resulting in three supercells, Pt(111)(a$\times$a$\times$b), where a indicates the number of replicas of the orthorhombic unit cell for Pt(111) surface, and b the number of layers in the slab. The three configurations considered are (2$\times$2$\times$3), (2$\times$2$\times$5), and (6$\times$6$\times$3). To model a simple interface, we adsorbed an ice-like water bilayer on the left surface of the Pt(111)(6$\times$6$\times$3) supercell while the right slab served as counter electrode. Both the H-down and H-up configurations \cite{Ogasawara2002, Bjorneholm2016} were considered. Our models consist of 24 water molecules: 12 form the $\mathrm{1^{st}}$ layer of chemisorbed molecules, while the remaining 12 form the $\mathrm{2^{nd}}$ layer. Additionally, we created low-coverage systems by removing 2 and 4 water molecules from the $\mathrm{1^{st}}$ water layer in both H-up and H-down models. For these models, the separation between the two plates was increased to 20 $\mathrm{\AA}$.\\
\begin{figure}[t]
    \centering
    \includegraphics[width=\columnwidth]{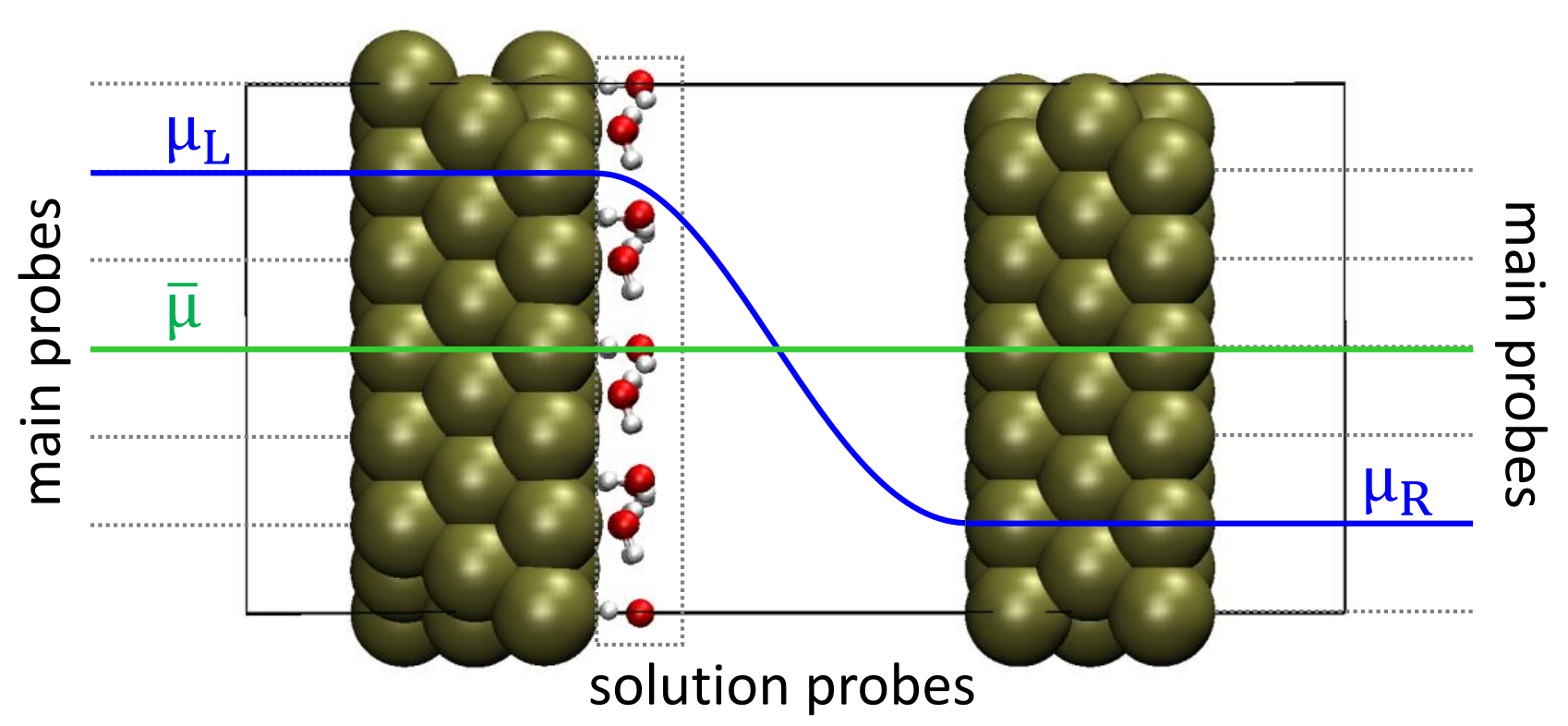}
    \caption{Representation of the parallel-plate capacitor models for HP-DFT calculations: two sets of main probes are coupled to the outer layers of each plate imposing $\mu_L$ and $\mu_R$ while solution probes are coupled to the atoms of the $\mathrm{H_2O}$ molecules. }
    \label{Fig:capacitors_models}
\end{figure}
The QUICKSTEP module employs a mixed Gaussian and plane waves (GPW) approach \cite{Lippert2010,Hutter2005_QS} to perform electronic structure calculations. In all our calculations the Perdew-Burke-Ernzerhof (PBE) exchange-correlation functional \cite{Perdew1996} was used in combination with the Grimme's D3 correction for dispersion interactions \cite{Grimme2010}. Goedeker-Teter-Hutter (GTH) pseudopotentials \cite{Goedecker1996} were used to describe the core electrons. TZVP-MOLOPT basis sets \cite{Hutter2007_MOLOPT} were used to expand the Kohn-Sham orbitals. For the Pt(111)(2$\times$2$\times$3) and Pt(111) (2$\times$2$\times$5) systems an 8x8x1 Monkhorst-Pack grid of k-points \cite{Monkhorst1976} was employed to sample the Brillouin zone. For the Pt(111)(6$\times$6$\times$3) system calculations were performed at the $\Gamma$-point. A conventional cubic cell with a lattice parameter of 3.97 $\mathrm{\AA}$ was employed to build the slabs. Periodic boundary conditions were applied only in the directions parallel to the surfaces of the plates (x and y), while in the direction perpendicular to them (z) an implicit Poisson solver was used \cite{Bani-Hashemian2016}. Threshold parameters for the SCF and force convergence were set at $1.0 \times 10^{-7}\ \mathrm{E_h}$ and $1.0 \times 10^{-3}\ \mathrm{E_h/a_0}$ respectively. The overall error associated with the total energy calculated with these parameters is $\pm 7.35\ \mathrm{E_h/cell}$. \\
For the HP-DFT calculations, two sets of main probes were coupled with the outermost layers of the capacitors, as depicted in Figure (\ref{Fig:capacitors_models}). The applied electrochemical potentials satisfied $\Delta \mu_L = -\Delta \mu_R$ to give $\mu_R = \overline{\mu} + \Delta\mu_R$ and $\mu_L = \overline{\mu} + \Delta\mu_L$. The resulting electrochemical potential difference across the system is $\Delta \mu = |\mu_L - \mu_R| $. In systems with water, solution probes were coupled with each atom of the molecules, as depicted in Figure (\ref{Fig:capacitors_models}). \\
The use of the HP-DFT formalism in calculations did not exhibit a significant deceleration of the SCF cycle in comparison to conventional DFT algorithms. Further information regarding the code's efficiency is available in Section 1 of the Supporting Material. The HP-DFT modules are available online at the \href{https://wiki.ch.ic.ac.uk/wiki/index.php?title=Running_a_HP-DFT_calculation_with_CP2K}{ICL Nano Electrochemistry Group's wiki}.

%% file: section_3.tex
\section{Results and discussion}
\begin{figure}[b]
    \centering
    \includegraphics[width=\columnwidth]{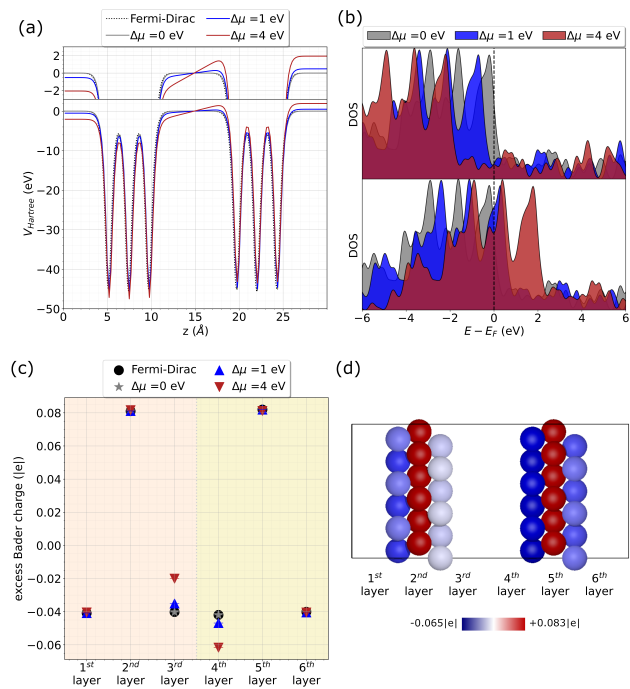}
    \caption{Pt(111)(6$\times$6$\times$3) system: (a) Average Hartree potential along the direction perpendicular to the surfaces ($\overline{\mu}$ is set to 0); (b)  PDOS of the atoms of the plates' inner layers ($\overline{\mu}$ is set to 0); (c) Average Bader charge per atom in each layer of the plates; (d) Bader charges for the Pt(111)(6$\times$6$\times$3) at $\Delta \mu = 4\ \mathrm{eV}$ }
    \label{Fig:Pt(111)(6x6x3)}
\end{figure}
\subsection{Capturing the physics of a parallel plate capacitor using HP-DFT} \label{Section:Clean_Surfaces }
\begin{table*}
    \centering
    \renewcommand{\arraystretch}{1.75} 
    \begin{tabular}{|>{\centering}m{0.9in} |>{\centering}m{0.9in} |>{\centering}m{0.9in} |>{\centering}m{0.9in} |>{\centering}m{0.9in} |>{\centering}m{0.9in} |>{\centering\arraybackslash}m{0.9in}|}    
        \hline 
        & \multicolumn{2}{c|}{\textbf{Pt(111)(2$\times$2$\times$3)}} & \multicolumn{2}{c|}{\textbf{Pt(111)(2$\times$2$\times$5)}} & \multicolumn{2}{c|}{\textbf{Pt(111)(6$\times$6$\times$3)}} \\
        \hline
        $\Delta \mu$ (eV) & q (C) & C (F) & q (C) & C (F) & q (C) & C (F) \\
        \hline
        1.00 &  $3.36\times 10^{-21}$ & $3.36 \times 10^{-21}$ &  $3.36\times 10^{-21}$ & $3.36\times 10^{-21}$ &  $3.04\times 10^{-20}$ & $3.04\times 10^{-20}$\\
        4.00 &  $1.35\times 10^{-20}$ & $3.36\times 10^{-21}$ &  $1.35\times 10^{-20}$ & $3.37\times 10^{-21}$ &  $1.22\times 10^{-19}$ & $3.04\times 10^{-20}$ \\
        \hline
        Theoretical* & \multicolumn{4}{c|}{$C_{(2\times2)surface}=2.99 \times 10^{-21}\ \mathrm{F}$ } & \multicolumn{2}{c|}{$C_{(6\times6)surface}=2.69 \times 10^{-20}\ \mathrm{F} $} \\
        \hline
    \end{tabular}
    \captionof{table}{Bader charges, q (C), and capacitance, C(F), at different values of applied $\Delta \mu$ for each model system. *values of capacitance calculated using the parallel-plate capacitor formula. }
    \label{Table:capacitance}
\end{table*}
In this section, we demonstrate the capability of the HP-DFT methodology to capture the properties of parallel-plate capacitor models, establishing its suitability to describe open-boundaries systems at fixed potential. \\
Firstly, we show that the imposed $\Delta \mu$ is reproduced and kept constant during the simulation. We applied an electrochemical potential difference between the two plates; values of $\Delta \mu = 1\ \mathrm{eV}$ and $\Delta \mu = 4\ \mathrm{eV}$ were considered. The average Hartree potential energy profile along the direction perpendicular to surfaces (shown in Figure (\ref{Fig:Pt(111)(6x6x3)}b)) reveals the existence of two local Fermi-like levels corresponding to $\Delta \mu_L$ and $\Delta \mu_R$. The Projected Density of States (PDOS) on the left and right plate (see Figure (\ref{Fig:Pt(111)(6x6x3)}c)) also shows a shift by $|\Delta \mu_{L}|$ towards lower energies and by $|\Delta \mu_{R}|$ towards higher energies respectively ($\overline{\mu}$ is set to zero in both cases). \\
Subsequently, we studied the charge distribution in the systems using Bader analysis \cite{Wang2009}. Figure (\ref{Fig:Pt(111)(6x6x3)}a) shows the average excess Bader charge per atom (defined as the difference between the calculated Bader charge and the atom valence) in each layer of the plates. Only the charge on the internal surfaces of the capacitor ($\mathrm{3^{rd}}$ and $\mathrm{4^{th}}$ layers) is modified with respect to the zero bias case as a consequence of the increase in $\Delta \mu$. In particular, the surface charge on the left slab becomes more positive while the surface charge on the right slab becomes more negative.\\ 
Finally, we calculated the capacitance for each of our models by evaluating $C=\sfrac{Q}{\Delta \mu}$, where $Q$ is the overall charge present on the plates. The values of capacitance thus calculated, reported in Table (\ref{Table:capacitance}), are in good agreement with those calculated using the well-known formula $C = \epsilon_0 \times (\sfrac{A}{d})$ (where $\epsilon_0$ is the vacuum permittivity, $A$ is the plates' surface area and $d$ is their separation). Our results also confirm that the capacitance only depends on surface geometry, yielding the same value of capacitance regardless of the applied $\Delta \mu$ or slab thickness, while it varies when the surface area of the plates changes. 
\subsection{The response of a water bilayer to the change of electrode potential} \label{Section:water_bilayer}
In this section, we analyse the response of a simple water bilayer/platinum interface to the application of an electrochemical potential and we show that the HP-DFT methodology manages to correctly reproduce key behaviours of water adsorption without the need for large systems and computationally costly calculations. \\
Although water adsorption is not an electrochemical step, as it doesn't involve an electron-coupled-proton transfer process, we find that the surface polarization induced by the applied potential has a significant impact on water coverage, structure and adsorption energy. This is in contrast with the N\o rskov assumption that only electrochemical steps in a reaction respond to changes in the electrode potential, while chemical steps are unaffected \cite{Norksov2004}. Moreover, we show that the ideal water bilayer model does not capture the adsorption and coverage behaviour of water over electrochemical interfaces under bias.\\
\subsubsection{Water Coverage and Charge redistribution} 
\begin{figure}
    \centering
    \includegraphics[width=\columnwidth]{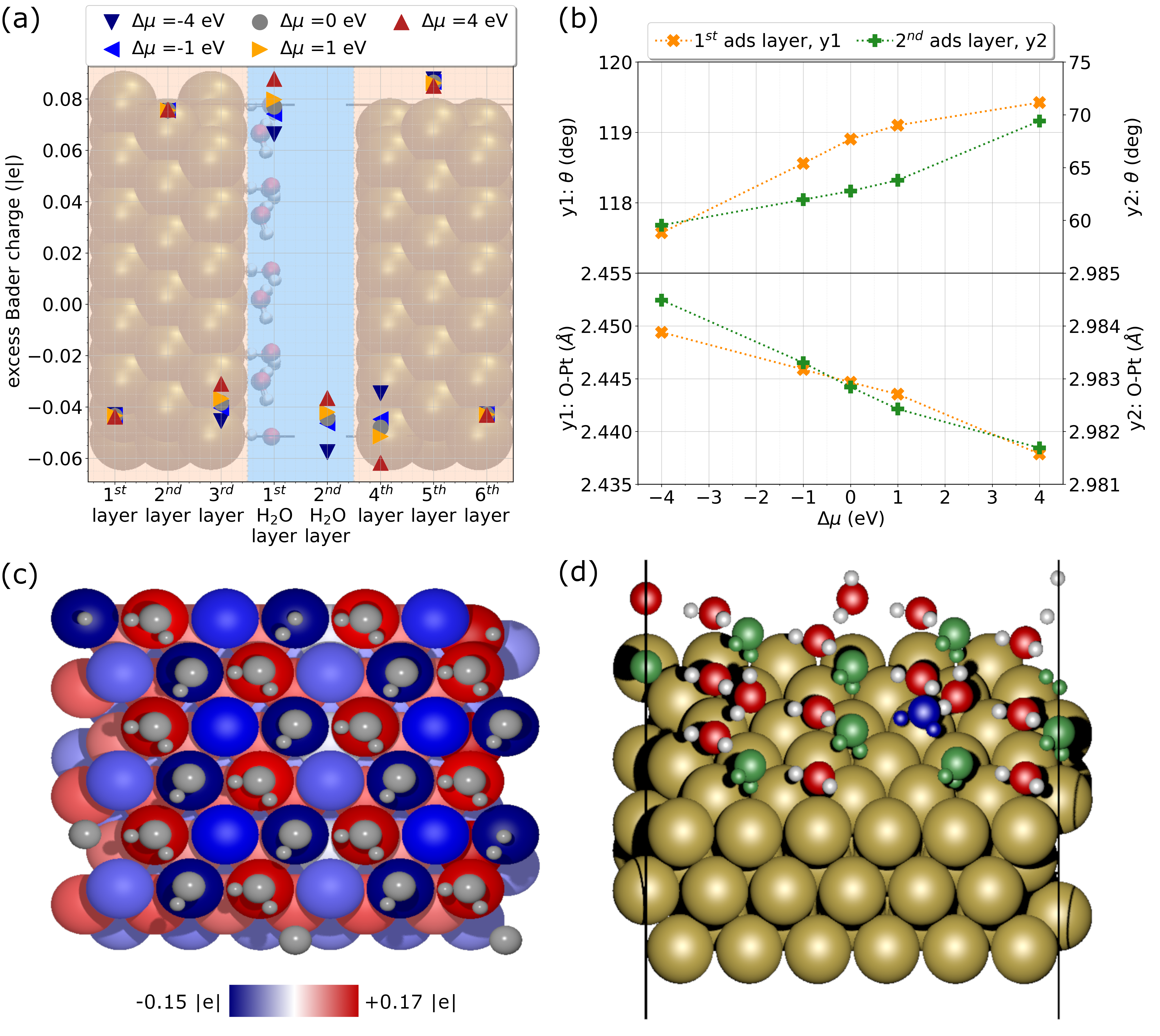}
    \caption{Pt(111)(6$\times$6$\times$3) system plus water bilayer: (a) average excess Bader charges of each atom of the plates and on each molecule of the bilayer, H-down configuration; (b) average $\theta$ and O-Pt distance for the molecules in the $\mathrm{1^{st}}$ (orange) and $\mathrm{2^{nd}}$ (green) adsorption layer, H-down configuration; (c) excess Bader charges on the left slab at $\Delta \mu = 4\ \mathrm{eV}$, H-down configurations; (d) Pt(111)(6$\times$6$\times$3) system plus H-up water bilayer at $\Delta \mu = -4\ \mathrm{eV}$: the green coloured molecules flipped from H-up to H-down configuration while the blue coloured one desorbed from the surface.}
    \label{Fig:Pt(111)(6x6x3)_bilayer}
\end{figure}

We studied water coverage and charge redistribution when $\Delta \mu$ goes from $-4\ \mathrm{eV}$ to $4\ \mathrm{eV}$. To this end, we calculated the adsorption energies as a function of $\Delta \mu$ for a series of structures with different coverages. The free energies of adsorption, $\Delta G_{ads}(\Delta \mu)$, were calculated according to \cite{Steinmann2020}:
\begin{equation}
\begin{split}
    \Delta G_{ads}&(\Delta \mu) = \frac{1}{{n_{H_2O}} } \times [E_\mathrm{system}(\Delta \mu) - E_\mathrm{slab}(\Delta \mu)  \\ & - (n_\mathrm{H_2O}\times G_\mathrm{H_2O}\mathrm{(gas)} - q_\mathrm{H_2O}(\Delta \mu)\times \Delta \mu_{L})]
    \label{Math:Gads}
\end{split}
\end{equation}
where $E_\mathrm{system}(\Delta \mu)$ is the total energy of the capacitor+water system at $\Delta \mu$, $E_\mathrm{slab}$ is the energy of the slabs without water at $\Delta \mu$ and $n_\mathrm{H_2O}$ is the number of adsorbed water molecules. $G_\mathrm{H_2O}\mathrm{(gas)}$ is the free energy of the water molecule in gas phase calculated as $G_\mathrm{H_2O}\mathrm{(gas)} = E_\mathrm{H_2O}\mathrm{(gas)} - TS $, where $E_\mathrm{H_2O}\mathrm{(gas)}$ is the energy of the molecule in gas phase obtained from the DFT calculation and $TS$ is the entropic contribution (this term was taken from ref \cite{Thrush}). Lastly, $q_\mathrm{H_2O}(\Delta\mu)$ is the charge transferred from the adsorbed water molecules to the surface and $\Delta \mu_L$ is the electrochemical potential of the slabs the bilayer is adsorbed on.\\ 
\begin{figure}[b]
    \centering
    \includegraphics[width=0.9\columnwidth]{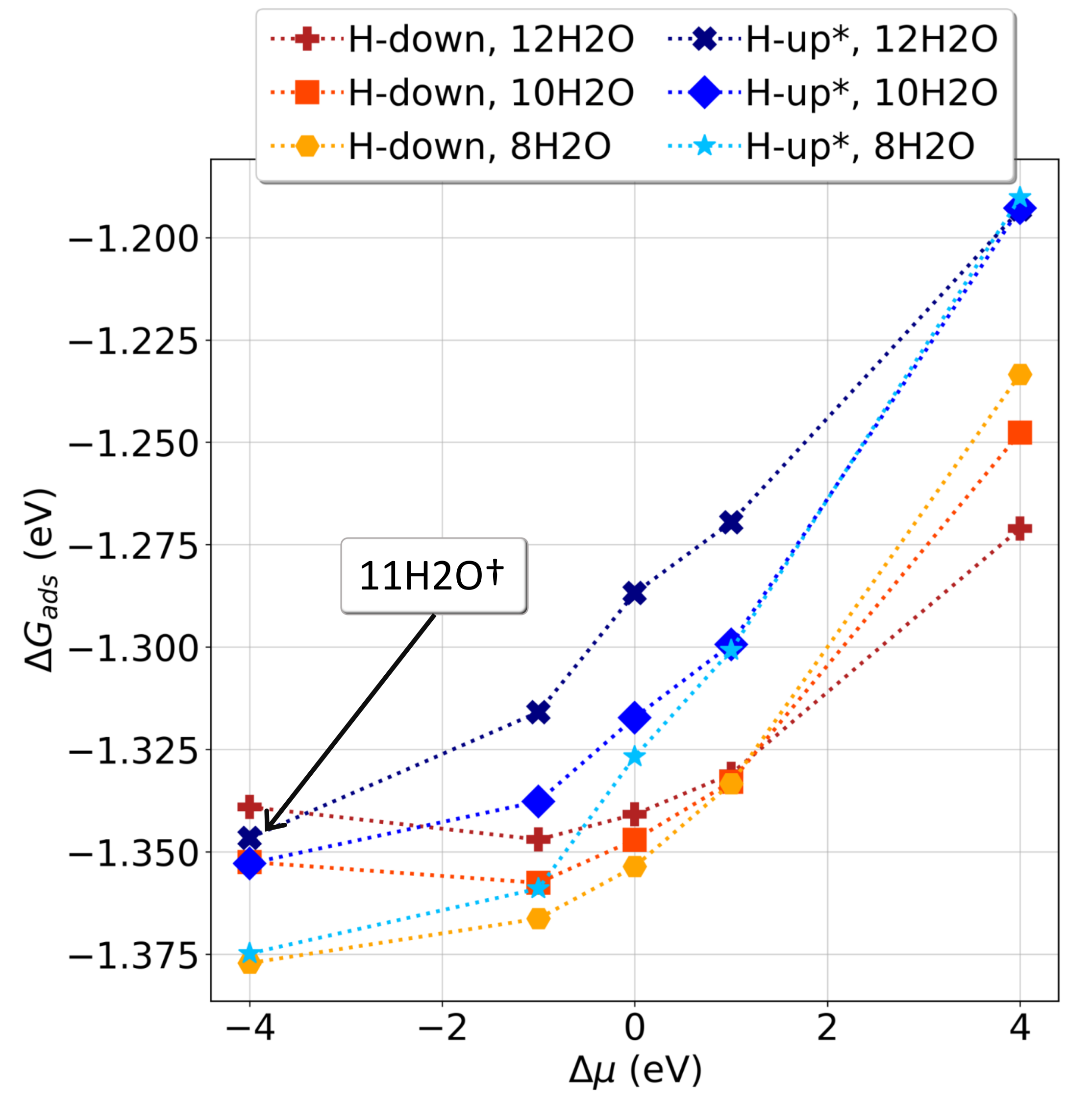}
    \caption{Free energy of adsorption for the water bilayer plotted as a function of the potential at different coverages: we compare both H-up and H-down configurations for cases with 12$\mathrm{H_2O}$, 10$\mathrm{H_2O}$, and 8$\mathrm{H_2O}$ chemisorbed water molecules. * The initial bilayer configuration is H-up, but several molecules flip to H-down at negative $\Delta \mu$. $^{\dag}$ One molecule desorbed from the $\mathrm{1^{st}}$ water layer.}
    \label{Fig:Pt(111)(6x6x3)_bilayer_Gads}
\end{figure} 
The results presented in Figure (\ref{Fig:Pt(111)(6x6x3)_bilayer_Gads}) show that, as the potential increases, the water coverage also increases. At $\Delta \mu = 0\ \mathrm{eV}$ and below the most stable configuration is the low coverage H-down bilayer, with only 8 chemisorbed $\mathrm{H_2O}$ molecules. The full coverage water bilayer becomes the most stable structure only at the very high $\Delta \mu = 4\ \mathrm{eV}$. At all potentials, the H-down structure remains the most stable configuration, with the exception of $\Delta \mu = -4\ \mathrm{eV}$. In this case, one of the chemisorbed molecules in the $\mathrm{1^{st}}$ layer spontaneously desorbed from the surface during geometry optimization, effectively reducing water coverage (blue-coloured molecule in Figure (\ref{Fig:Pt(111)(6x6x3)_bilayer}d)). \\
The stabilisation of low water coverage structures at low potential can be explained in terms of the charge redistribution when the potential changes. As $\Delta \mu$ becomes progressively more positive the charge on the left plate becomes overall more positive, accumulating as expected only on the internal surface of the capacitor, as shown in Figure (\ref{Fig:Pt(111)(6x6x3)_bilayer}a). Our data show that chemisorbed water molecules in the water bilayer transfer electrons to the metal surface, becoming positively charged, with an average charge between 0.06 and 0.09 $\mathrm{|e|}$, depending on coverage and potential (see Table (S1) of the Supporting Material). The metal atoms to which they bind also have a net positive charge; the negative charge from the O-Pt bond spills over the neighbouring metal surface atoms, causing them to become negatively charged, as shown in Figure (\ref{Fig:Pt(111)(6x6x3)_bilayer}c). The presence of these negative Pt atoms destabilises the ice-like bilayer structure even at $\Delta \mu = 0\ \mathrm{eV}$ (see Table (S2) of the supporting Material). When the potential becomes more positive, the coverage of positively charged water molecules can increase but it saturates when additional negative charge transfer from water to the surface becomes unfeasible. At low potentials, the system cannot accommodate the positive bilayer, leading to the observed desorption. \\
The observations discussed in this section all align with recent AIMD studies \cite{Cheng2018,Khatib2021,Neugebauer2021,Darby2022} which demonstrate that at zero bias the water coverage on the Pt electrode is lower than that modelled in the standard bilayer model. Our results also align with recent literature in showing that capacitive response of the interface to a positive potential is primarily driven by the increase in surface coverage of positively charged water molecules. It is worth noting that these studies use exceptionally large and realistic models to simulate explicitly the polarization of the interfacial water layer when a potential is applied. However, this level of realism is achieved at the expense of substantial computational resources, while our results were obtained using significantly smaller systems. 
\subsubsection{Water reorientation}
Surface polarisation is also compensated by structural changes in both water layers, as shown in Figure (\ref{Fig:Pt(111)(6x6x3)_bilayer}b). As $\Delta \mu$ becomes more positive, the O-Pt distance decreases and $\theta$ (the angle between the normal to the surface and the water dipole) increases for both the $\mathrm{1^{st}}$ and $\mathrm{2^{nd}}$ water layers. In the $\mathrm{1^{st}}$ layer, water molecules slightly reorient their H atoms towards or away from the surface at negative or positive potentials respectively. The molecules of the $\mathrm{2^{nd}}$ layer present the same behaviour, but respond more freely to the change in surface charge and the observed reorientation is larger. \\
The analysis of interfacial charge redistribution also helps to explain water reorientation (see Figure (\ref{Fig:Pt(111)(6x6x3)_bilayer}c)). In the H-down structure, the attractive Coulombic interaction with the partially positive H stabilises the negative charge on the Pt atoms. Indeed, the molecules of the $\mathrm{2^{nd}}$ adsorption layer of the H-down configuration are closer to the surface (cf. Figure (\ref{Fig:Pt(111)(6x6x3)_bilayer}b) and Figure (S2b) of the Supporting Material). Conversely, in the H-up structure the partially negative oxygen atoms of the $\mathrm{2^{nd}}$ water layer point towards the surface, leading to Coulombic repulsion and a reduction in the negative charge on the Pt atom (lighter shades of blue in Table (S2) of the Supporting Material). \\
The trends discussed are clearly observable only for the more stable H-down configuration. However consistent observations arise also from the analysis of the H-up system, although a clear geometrical reorganisation trend is only seen for positive $\Delta \mu$ values, as can be observed from Figure (S2) and Table (S2) of the Supporting Material. This lack of monotonic trend in angles and distance is due to several molecules in the $\mathrm{2^{nd}}$ layer flipping from an H-up to an H-down configuration at negative $\Delta \mu$. For the full bilayer structures the number of molecules flipping at $\Delta \mu = -1\ \mathrm{eV}$ and $-4\ \mathrm{eV}$ is, respectively, 2 and 9 out of 12 molecules in the $\mathrm{2^{nd}}$ layer (green molecules in Figure (\ref{Fig:Pt(111)(6x6x3)_bilayer}d)). We observed this process for the low-coverage systems as well (Table (S2) of the Supporting Material). This so-called flip-flop process of the water molecules at charged interfaces has been experimentally documented \cite{Nihonyanagi2009,Nihonyanagi2013} and can be rationalized in terms of the response of the water molecule to the increasingly negative surface charge: the H-down configuration brings the positive part of the dipole closer to the increasingly negative surface and stabilizes the molecule. \\ 

%% file: conclusions.tex
\section{Conclusions}
We present a novel and efficient open-boundary formalism, HP-DFT, for modelling electrochemical systems. It combines the established Hairy Probes (HP) formalism with DFT in the canonical ensemble, offering computational efficiency comparable to standard DFT. \\
The HP-DFT formalism accurately reproduces parallel-plate capacitor models' properties, demonstrating its suitability to describe open-boundary systems at fixed potential. The formalism successfully reproduces the imposed $\Delta \mu$ and keeps it constant during the simulations. Furthermore, the physics of the parallel-plate capacitor is correctly described and the capacitance obtained is in agreement with the theoretically evaluated one. \\
When applied to a water bilayer/platinum interface, HP-DFT successfully captures the key features of water adsorption on a charged metal plate, correctly describing charge redistribution and water reorientation. Our study finds that coverage, interface structure and adsorption energies are strongly influenced by the applied potential, despite water adsorption not being an electrochemical process. This is in contrast with N\o rskov's proposed model. Additionally, we find that the water bilayer doesn't accurately represent the adsorption behaviour of a full water layer at electrochemical interfaces, as systems with lower coverage are energetically more stable even at zero bias. The results we found align with previous AIMD studies. This highlights the advantages of using a methodology, HP-DFT, to directly control the electrode potential, as it can capture polarization response using small models and fine-tune the analysis of potential effects. Considering how lightweight the code itself is, the HP-DFT formalism proves an excellent candidate to model EIs by means of AIMD calculations with an open-boundary description of the electrons. 

%% file: acknowledgements.tex
\section{Acknowledgements}
 This work was supported by the Engineering and Physical Sciences Research Council (grant EP/P033555/1). We would like to acknowledge the Thomas Young Centre under grant number TYC-101. We also gratefully acknowledge the use of the High-Performance Computers at Imperial College London, provided by Imperial College Research Computing Service DOI: 10.14469/hpc/2232, and the computing resources provided by STFC Scientific Computing Department’s SCARF cluster. This research also used ARCHER2 UK National Supercomputing Service (https://www.archer2.ac.uk), via our membership of the UK's HEC Materials Chemistry Consortium, which is funded by EPSRC (EP/R029431 and EP/X035859). We are thankful for the computational support offered by Dr Alin Marin Elena and for the insightful discussions with our colleague and fellow researcher, Dr Marcella Iannuzzi, who greatly helped us understand the inner workings of the CP2K code. 